\documentclass[12pt]{article}

\title{\textbf{Secure Key Distribution via Pre- and Post-Selected Quantum 
States}}
\author{Jeffrey Bub\thanks{Email address: jbub@carnap.umd.edu} \\ 
\small \textit{Philosophy Department, University of Maryland, College Park, 
MD 20742, USA}}
\date{}

\normalsize
\begin{document}
\maketitle

\begin{abstract}
A quantum key distribution scheme whose security depends on the 
features of pre- and post-selected quantum states is described.
\end{abstract}  

\bigskip


\section{Introduction}

A wide variety of quantum key distribution schemes have been 
proposed, following the original Bennett and 
Brassard protocol~\cite{BB84}. Ekert~\cite{Ekert} has described a scheme 
in which two 
parties, Alice and Bob, create a shared random key by performing spin 
measurements on 
pairs of spin-$\frac{1}{2}$ particles in the singlet state. The 
particle pairs are emitted by a source towards Alice and Bob, 
who each measure spin along three different directions, chosen 
randomly and independently for each pair. After a sequence of 
measurements on an appropriate number of pairs, Alice and Bob 
announce the directions of their measurements publicly and divide the 
measurements into two groups: those in which they measured the spin 
in different directions, and those in which they measured the spin in 
the same direction. They publicly reveal the outcomes of the first group of 
measurements and use these to 
check that the singlet states have not been disturbed by an 
eavesdropper, Eve. Essentially, they calculate a correlation 
coefficient: any attempt by 
Eve to monitor the particles will disturb the singlet state and 
result in a correlation coefficient 
that is bounded by Bell's inequality and is hence distinguishable 
from the correlation 
coefficient for the singlet state. If 
Alice and Bob are satisfied that no eavesdropping has occurred, they 
use the second group of (oppositely correlated) measurement outcomes as 
the raw key.

The Ekert scheme solves the key distribution problem as well as the 
key storage problem, because there is no information in the singlets 
before Alice and Bob perform their measurements and communicate 
classically to establish the key. 
The scheme proposed here also involves entangled states, but 
the test for eavesdropping is different. Instead of a statistical test 
based on Bell's theorem, the test exploits conditional statements about 
measurement outcomes generated by pre- and 
post-selected quantum states.

\section{Pre- and post-selected quantum states}

The peculiar features of pre- and post-selected quantum states 
were first pointed out by 
Aharonov, Bergmann, and 
Lebowitz~\cite{ABL}. If (1) Alice prepares a system in a 
certain state $|\mbox{pre}\rangle$ 
at time $t_{1}$, (2) Bob measures some observable $Q$ on the system 
at time $t_{2}$, and (3) Alice measures an observable of which 
$|\mbox{post}\rangle$ is an eigenstate at time $t_{3}$, 
and post-selects for $|\mbox{post}\rangle$, then Alice can assign 
probabilities to the outcomes of Bob's $Q$-measurement at $t_{2}$, 
conditional on the states $|\mbox{pre}\rangle$ and $|\mbox{post}\rangle$ at times 
$t_{1}$ and $t_{3}$, respectively, as follows \cite{ABL,VAA}:
\begin{equation}
    \mbox{prob}(q_{k}) =
    \frac{|\langle \mbox{pre}|P_{k}| \mbox{post}\rangle|^{2}}
    {\sum_{i} |\langle \mbox{pre} |P_{i}|\mbox{post}\rangle|^{2}}
    \label{eq:ABL}
\end{equation}
where $P_{i}$ is the projection operator onto the $i$'th eigenspace 
of $Q$. Notice that (\ref{eq:ABL})---referred to as the `ABL-rule' 
(Aharonov-Bergmann-Lebowitz rule) in the following---is 
time-symmetric, in the sense that the states $|\mbox{pre}\rangle$ and 
$|\mbox{post}\rangle$ can be interchanged.

If $Q$ is unknown to Alice, she can use the ABL-rule to assign 
probabilities to the outcomes of various hypothetical 
$Q$-measurements. The interesting peculiarity of the ABL-rule, by 
contrast with the usual Born rule for pre-selected states, is that it 
is possible---for an appropriate choice of observables $Q$, $Q'$, 
\ldots, and states $|\mbox{pre}\rangle$ and $|\mbox{post}\rangle$---to 
assign unit probability to the outcomes of a set of mutually 
\textit{noncommuting} observables. That is, Alice can be in a 
position to assert a conjunction of conditional statements of the 
form: `If Bob measured $Q$, then the outcome must have been $q_{i}$, 
with certainty, and if Bob measured $Q'$, then the outcome must have been 
$q'_{j}$, with certainty, \ldots,' where $Q, Q', \ldots$ are mutually 
noncommuting observables. Since Bob could only have measured at most 
one of these noncommuting observables, Alice's conditional information 
does not, of course, contradict quantum mechanics: she only knows the 
eigenvalue $q_{i}$ of an observable $Q$ if she knows that Bob in fact 
measured $Q$. 

Vaidman, Aharonov, and Albert~\cite{VAA} discuss a case of this sort, 
where the outcome of a 
measurement of any of the three spin components $\sigma_{x}$, 
$\sigma_{y}$, $\sigma_{z}$ of a spin-$\frac{1}{2}$ particle can be 
inferred from an appropriate pre- and post-selection. Alice prepares 
the Bell state:
\begin{equation}
    |\mbox{pre}\rangle = 
    \frac{1}{\sqrt{2}}(|\uparrow_{z}\rangle_{A}|\uparrow_{z}\rangle_{C} + 
    |\downarrow_{z}\rangle_{A}|\downarrow_{z}\rangle_{C}
    \label{eq:Bell}
\end{equation}
where $|\uparrow_{z}\rangle$ and $|\downarrow_{z}\rangle$  
denote the $\sigma_{z}$-eigenstates. Alice sends one of the 
particles---the channel particle, denoted by the subscript $C$---to Bob 
and keeps the ancilla, denoted by $A$. Bob measures either 
$\sigma_{x}$, or $\sigma_{y}$, or $\sigma_{z}$ on the channel 
particle and returns the channel particle to Alice. Alice then 
measures an observable $R$ on the pair of particles, where $R$ has 
the eigenstates:
\begin{eqnarray}
    |r_{1}\rangle & = & 
    \frac{1}{\sqrt{2}}|\uparrow_{z}\rangle|\uparrow_{z}\rangle + 
    \frac{1}{2}(|\uparrow_{z}\rangle|\downarrow_{z}\rangle 
    e^{i\pi/4} + |\downarrow_{z}\rangle|\uparrow_{z}\rangle 
    e^{-i\pi/4}) \\
    |r_{2}\rangle & = & 
    \frac{1}{\sqrt{2}}|\uparrow_{z}\rangle|\uparrow_{z}\rangle - 
    \frac{1}{2}(|\uparrow_{z}\rangle|\downarrow_{z}\rangle 
    e^{i\pi/4} + |\downarrow_{z}\rangle|\uparrow_{z}\rangle 
    e^{-i\pi/4}) \\
    |r_{3}\rangle & = & 
    \frac{1}{\sqrt{2}}|\downarrow_{z}\rangle|\downarrow_{z}\rangle + 
    \frac{1}{2}(|\uparrow_{z}\rangle|\downarrow_{z}\rangle 
    e^{-i\pi/4} + |\downarrow_{z}\rangle|\uparrow_{z}\rangle 
    e^{i\pi/4}) \\
    |r_{4}\rangle & = & 
    \frac{1}{\sqrt{2}}|\downarrow_{z}\rangle|\downarrow_{z}\rangle - 
    \frac{1}{2}(|\uparrow_{z}\rangle|\downarrow_{z}\rangle 
    e^{-i\pi/4} + |\downarrow_{z}\rangle|\uparrow_{z}\rangle 
    e^{i\pi/4})
\end{eqnarray}

Note that:
\begin{eqnarray}
    |\mbox{pre}\rangle & = & 
    \frac{1}{\sqrt{2}}(|\uparrow_{z}\rangle|\uparrow_{z}\rangle + 
    |\downarrow_{z}\rangle|\downarrow_{z}\rangle \\ \label{eq:R1}
                       & = & 
    \frac{1}{\sqrt{2}}(|\uparrow_{x}\rangle|\uparrow_{x}\rangle + 
    |\downarrow_{x}\rangle|\downarrow_{x}\rangle \\
                       & = & 
    \frac{1}{\sqrt{2}}(|\uparrow_{y}\rangle|\downarrow_{y}\rangle + 
    |\downarrow_{y}\rangle|\uparrow_{y}\rangle \\
                       & = & 
    \frac{1}{2}(|r_{1}\rangle + |r_{2}\rangle + |r_{3}\rangle + 
    |r_{4}\rangle) '\label{eq:R4}
\end{eqnarray}
In Eqs. (\ref{eq:R1})--(\ref{eq:R4}) and in the following, the 
subscripts $A$ and $C$ appearing in Eq. (\ref{eq:Bell}) are implicit in the 
tensor product notation. 
Eqs. (\ref{eq:R1})--(\ref{eq:R4}) correspond to Eq. (2) of 
\cite{VAA} or Eq. (54) of \cite{Metzger}.
    
    Alice can now assign values to the outcomes of Bob's spin measurements  
via the ABL-rule,
whether Bob measured $\sigma_{x}$, $\sigma_{y}$, or $\sigma_{z}$,
based on the post-selections $|r_{1}\rangle$, $|r_{2}\rangle$, 
$|r_{3}\rangle$, or $|r_{4}\rangle$, according to Table~\ref{table:xyz}  
(where 0 represents the outcome $\uparrow$ and 1 represents the 
outcome $\downarrow$) \cite{VAA}:
\begin{table}[ht]
    \begin{center}
  $
    \begin{array}{r|ccc}
	      & \sigma_{x} & \sigma_{y} & \sigma_{z} \\ \hline
        r_{1} & 0 & 0 & 0 \\ 
        r_{2} & 1 & 1 & 0 \\ 
        r_{3} & 0 & 1 & 1 \\ 
        r_{4} & 1 & 0 & 1 
     \end{array}
  $
  \end{center}
    \caption{\protect $\sigma_{x}$, \protect $\sigma_{y}$, \protect 
    $\sigma_{z}$ measurement outcomes correlated with eigenvalues of R}
    \label{table:xyz}
\end{table}

\section{The key distribution protocol}

\begin{figure}

\begin{picture}(350,100)
    
\put(0,70){\line(1,0){25}}
\put(25,60){\framebox(20,20){\textit{H}}}
\put(45,70){\line(1,0){50}}
\put(70,70){\circle*{7}}
\put(95,60){\framebox(30,20){\textit{Bob}}}
\put(125,70){\line(1,0){190}}
\put(150,70){\circle*{7}}
\put(152,80){$\pi$}
\put(175,70){\circle{7}}
\put(200,70){\circle*{7}}
\put(230,70){\circle*{7}}
\put(260,70){\circle*{7}}
\put(290,70){\circle*{7}}
\put(285,80){$-\frac{3\pi}{4}$}
\put(315,60){\framebox(20,20){\textit{H}}}
\put(335,70){\line(1,0){25}}

\put(0,20){\line(1,0){220}}
\put(70,20){\circle{7}}
\put(175,20){\circle*{7}}
\put(200,20){\circle*{7}}
\put(220,10){\framebox(20,20){\textit{H}}}
\put(240,20){\line(1,0){120}}
\put(260,20){\circle*{7}}

\put(70,20){\line(0,1){50}}
\put(175,20){\line(0,1){50}}
\put(200,22){\line(0,1){50}}
\put(202,40){$\frac{\pi}{2}$}
\put(230,30){\line(0,1){40}}
\put(260,22){\line(0,1){50}}
\put(262,40){$\frac{\pi}{2}$}

\end{picture}

\caption{Quantum circuit for key distribution protocol}
    \label{fig}
\end{figure}
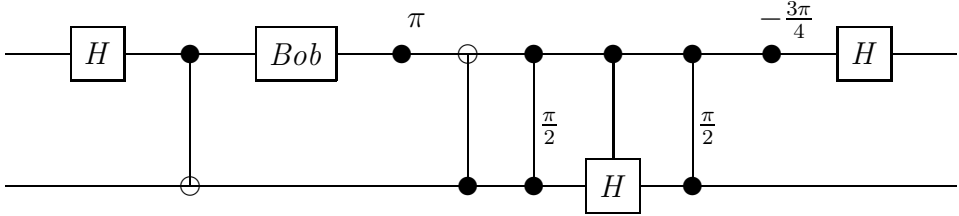

This case can be exploited to enable Alice and Bob to share a private 
random key in the following way: Alice prepares a certain number of 
copies (depending on the length of the key and the level of privacy 
desired) of the Bell state, Eq. (\ref{eq:Bell}).
She sends the channel particles to Bob in sequence and keeps the 
ancillas. Bob measures $\sigma_{x}$ or $\sigma_{z}$ randomly on the 
channel particles and returns the particles, in sequence, to Alice. 
Alice then measures the observable $R$ on the ancilla and channel 
pairs and divides the sequence into two subsequences: the 
subsequence $S_{14}$ 
for which she obtained the outcomes $r_{1}$ or $r_{4}$, and the 
subsequence $S_{23}$ for which she obtained the outcomes $r_{2}$ or 
$r_{3}$. The sequence of quantum operations can be implemented on a 
quantum circuit as in Fig.\ \ref{fig}  (see Eq. (46) of Metzger~\cite{Metzger}). 
In the present work, an ideal system without noise is assumed.

To check that the channel particles have not been monitored by Eve, 
Alice now publicly 
announces the indices of the subsequence $S_{23}$. As is 
evident from Table~\ref{table:xyz}, 
for this subsequence she can make conditional 
statements of the form: `For channel particle $i$, if $\sigma_{x}$ 
was measured, the outcome was 1 (0), and if $\sigma_{z}$ 
was measured, the outcome was 0 (1),' depending on whether the outcome of her 
$R$-measurement was $r_{2}$ or $r_{3}$. She announces these 
statements publicly. If one of these statements, for some index $i$, 
does not agree with Bob's records, Eve must have monitored the $i$'th 
channel particle. (Of course, agreement does not entail that the 
particle was \textit{not} monitored.)

For suppose Eve 
measures a different spin component observable than Bob on a channel particle 
and Alice subsequently obtains one of the eigenvalues $r_{2}$ or $r_{3}$ 
when she measures $R$. 
Bob's measurement outcome, either 0 or 1, 
will be compatible with just one of these eigenvalues, assuming no 
intervention by Eve. 
But after Eve's measurement, both of these eigenvalues will 
be 
possible outcomes of Alice's measurement. So Alice's retrodictions of 
Bob's measurement outcomes for the subsequence $S_{23}$ 
will not necessarily correspond to Bob's 
records. In fact, it is easy to see that if Eve measures $\sigma_{x}$ 
or $\sigma_{z}$ randomly on the channel particles, or if she measures 
a particular one of the observables $\sigma_{x}$, $\sigma_{y}$, or 
$\sigma_{z}$ on the channel particles (the same observable on each 
particle), the probability of detection in the subsequence $S_{23}$ 
is 3/8.   

In the subsequence $S_{14}$, the 0 and 1 outcomes of Bob's measurements 
correspond to the outcomes $r_{1}$ and $r_{4}$ of Alice's 
$R$-measurements. If, following their public communication about the 
subsequence $S_{23}$, Alice and Bob agree that there has been no 
monitoring of the 
channel particles by Eve, they use the subsequence $S_{14}$ to define 
a shared raw key.

Note that even a single disagreement between Alice's 
retrodictions and Bob's records is sufficient to reveal that the channel 
particles have been monitored by Eve. This differs from the 
eavesdropping test in the Ekert protocol. 
Note also that Eve only has access to the channel particles, not 
the particle pairs. So no strategy is possible in which Eve replaces all the 
channel particles with her own particles and 
entangles the original channel particles, treated as a single system, 
with an ancilla by some unitary transformation, and then delays any
measurements until after Alice and Bob have communicated publicly. 
There is no way that Eve can ensure agreement between Alice and Bob 
without having access to the particle pairs, or without information 
about Bob's measurements.

The key distribution protocol as outlined above solves the key 
distribution problem but not the key storage problem. If Bob actually 
makes the random choices, measures $\sigma_{x}$ or $\sigma_{z}$, and 
records definite outcomes for the spin measurements 
before Alice measures $R$, as required by the protocol, Bob's 
measurement records---stored as classical information---could in 
principle be 
copied by Eve without detection. In that case, 
Eve would know the raw key (which is 
contained in this information), following the public communication
between Alice and 
Bob to verify the integrity of the quantum communication channel. 

To solve the key storage problem, the protocol is modified in the 
following way: Instead of actually making the random choice for each 
channel particle,  
measuring one of the spin observables, and recording the outcome of the 
measurement, Bob keeps the random choices and the spin 
measurements `at the quantum level' until after Alice announces the 
indices of the subsequence $S_{23}$ of her $R$ measurements. To do 
this, Bob enlarges the Hilbert space by entangling the 
quantum state of the channel particle via a unitary transformation with 
the states of two ancilla 
particles that he introduces. One particle is associated with a 
Hilbert space spanned by two 
eigenstates, $|c_{\sigma(x)}\rangle$ and $|c_{\sigma(z)}\rangle$, of a choice
observable $C$. The 
other particle is associated with a 
Hilbert space spanned by two eigenstates, $|p_{\uparrow}\rangle$ and 
$|p_{\downarrow}\rangle$, of a pointer 
observable $P$. (See \cite{Lo}, 
footnote \textit{t}, or \cite{Bub} for details of how to implement the 
unitary transformation on the enlarged Hilbert space.) 

On the modified protocol (assuming the ability 
to store entangled states indefinitely), Alice and Bob share a large 
number of copies of an entangled 4-particle state. When they wish to establish a 
random key of a certain length, Alice measures $R$ on an appropriate 
number of particle pairs in her possession and announces the indices of the 
subsequence $S_{23}$. Before Alice announces the indices of the 
subsequence $S_{23}$, neither Alice nor Bob have stored any classical information. So there is nothing for Eve to copy.
After Alice announces the 
indices of the subsequence $S_{23}$, Bob measures the observables $D$ 
and $P$ on his ancillas
with these indices and announces 
the eigenvalue $|p_{\uparrow}\rangle$ or $|p_{\downarrow}\rangle$ as the
outcome of his $\sigma(x)$ or $\sigma(z)$ measurement, depending on 
the eigenvalue of $D$. If Alice and 
Bob decide that there has been no eavesdropping by Eve, Bob measures 
$D$ and $P$ on his ancillas in the 
subsequence $S_{14}$. It is easy 
to see that the ABL-rule applies in this case, just as it applies 
in the case where Bob actually makes the random choice and actually 
records definite outcomes of his $\sigma(x)$ or $\sigma(z)$ 
measurements before Alice measures $R$. (In 
fact, if the two cases were not equivalent for Alice---if Alice could 
tell from her $R$-measurements 
whether Bob had actually made the random choice and actually performed 
the spin measurements, or had merely implemented these actions 
`at the quantum level'---the difference could be 
exploited to signal superluminally.) 

There are clearly other possible ways of exploiting this case 
to implement a secure key distribution 
protocol (involving all three spin component observables, for 
example), but the principle is similar. It would seem worthwhile to 
consider whether other applications of pre- and post-selection might be applied 
as a tool in quantum cryptology.

\section*{Acknowledgements}
This work was partially supported by a University of Maryland 
General Research Board leave fellowship. Illuminating discussions with 
Gilles Brassard,  
Lev Vaidman, and especially Adrian Kent 
are acknowledged with thanks.


\begin{thebibliography}{99}
    
\bibitem{ABL} Y. Aharanov, P.G. Bergmann, and J.L Lebowitz, \textit{Phys. 
Rev. B} {\bf 134}, 1410--1416. Reprinted in J. A. Wheeler and W. H. Zurek (eds.),
\textit{Quantum Theory and 
Measurement}  (Princeton: Princeton 
University Press, 1983), pp. 680--686.
    
\bibitem{Bub} J. Bub, `The quantum bit commentment theorem,' quant-ph/007090. 
Forthcoming in \textit{Foundations of Physics}.
    
\bibitem{BB84} C.H. Bennett and G. Brassard, `Quantum cryptography: 
public key distribution and coin tossing,' in \textit{Proceedings of IEEE 
International Conference on Computers, Systems, and Signal 
Processing}, pp. 175--179. IEEE, 1984.  

\bibitem{Ekert} A. Ekert, Phys. Rev. Letters {\bf 67}, 
661 (1991).

\bibitem{Lo} H.-K. Lo, `Quantum cryptology,' 
in H.-K. Lo, S. Popescu, and T. Spiller (eds.),
\textit{Introduction to Quantum Computation and Information}
(Singapore: World Scientific, 1998). 

\bibitem{Metzger} S. Metzger, `Spin-measurement retrodiction revisited,'
quant-ph/0006115.

\bibitem{VAA} L. Vaidman, Y. Aharonov, and D.Z. Albert, 
Phys. Rev. Letters {\bf 58}, 1385 (1987).




\end{thebibliography}
\end{document}